# *BABAR* – A Community Web Site in an Organizational Setting

Ray Cowan
*Laboratory for Nuclear Science, M.I.T., Cambridge, MA 02139*

Yogesh Deshpande
*University of Western Sydney, Locked Bag 1797, Penrith South DC, NSW 1797, Australia*

Bebo White
*Stanford Linear Accelerator Center, 2575 Sand Hill Road, Menlo Park, CA*

For the *BABAR* Computing Group

The *BABAR* Web site was established in 1993 at the Stanford Linear Accelerator Center (SLAC) to support the *BABAR* experiment, to report its results, and to facilitate communication among its scientific and engineering collaborators, currently numbering about 600 individuals from 75 collaborating institutions in 10 countries. The *BABAR* Web site is, therefore, a community Web site. At the same time it is hosted at SLAC and funded by agencies that demand adherence to policies decided under different priorities. Additionally, the *BABAR* Web administrators deal with the problems that arise during the course of managing users, content, policies, standards, and changing technologies. Desired solutions to some of these problems may be incompatible with the overall administration of the SLAC Web sites and/or the SLAC policies and concerns. There are thus different perspectives of the same Web site and differing expectations in segments of the SLAC population which act as constraints and challenges in any review or re-engineering activities. Web Engineering, which post-dates the *BABAR* Web, has aimed to provide a comprehensive understanding of all aspects of Web development. This paper reports on the first part of a recent review of application of Web Engineering methods to the *BABAR* Web site, which has led to explicit user and information models of the *BABAR* community and how SLAC and the *BABAR* community relate and react to each other. The paper identifies the issues of a community Web site in a hierarchical, semi-governmental sector and formulates a strategy for periodic reviews of *BABAR* and similar sites. A separate paper reports on the findings of a user survey and selected interviews with users, along with their implications and recommendations for future

## 1. INTRODUCTION

The *BABAR* Web site was established in 1993 at SLAC to support the *BABAR* experiment, to report its results, and to facilitate communication among its scientific and engineering collaborators. The experiment currently numbers about 600 individuals from 75 collaborating institutions in 10 countries[1]. Viewed in this, its primary role, the *BABAR* Web site is a scientific and professional community Web site where the users expect to be treated in an open, collaborative and consultative manner. The users expect the site administrators to make the working environment as friendly and open as possible, to listen to the users' complaints, and to solve problems by instituting the necessary changes including adjustments for the changing technologies. The users expect minimum hierarchical constraints and maximum ease of use.

The *BABAR* Web site has evolved over the last 10 years and is now essential not only to the community, i.e. the collaborators, but also to funding agencies and educational institutions. The collaborators use many different platforms and technologies to interact with the Web site. The site is mirrored to insure its freshness for international collaborators. While the site's traffic does not approach that of commercial Web sites, the daily number of hits is high by high energy physics standards. It is, therefore, critical that the administrators create, follow and enforce standards to facilitate the growth and usage of the site. The first major review of the *BABAR* Web took place in 1998[1]. The present site, both in its public interface and internal organization, is a result of that review when the user community was widely consulted.

The current review, conducted in March and April 2003, arose out of formal and informal discussions over time that indicated some levels of dissatisfaction between the content providers and the site administrators, both at the *BABAR* and SLAC levels. The catalogue of informal suggestions and complaints could be broadly classified as:

a) Internal improvements to the existing *BABAR* Web site - searching capability, removal of outdated information, some restructuring of information architecture, and better maintenance procedures;

b) Additional capabilities - adoption of diverse and/or newer technologies, more freedom in creating Web pages, and new applications;

c) Stricter adherence to the SLAC procedures, in particular with respect to security concerns.

The *BABAR* Web site is not the only community site with some unique problems. SLAC supports additional Web sites for other high energy physics experiments. This is also true for other international research institutions. The current review may therefore be of interest to Web administrators at these other sites.

This paper is one of two, planned reports on the review of the *BABAR* Web. It charts the evolution of the *BABAR* Web site, its community, perceptions of the different interest-groups making up the whole community, and the complexity of the *BABAR* Web site's operations. A separate paper will report on the findings of the survey and interviews of the *BABAR* community, and on the current policies and strategies to deal with Web technologies, standards, tools, techniques and methods used.

The paper is organized as follows. Section 2 describes the *BABAR* Web environment, viz. its audience, hardware and software platforms and information content. Section 3 deals with the broader goals of a virtual community. Section 4 examines the *BABAR* Web in the context of the operations and needs of its virtual community. Section 5 brings out the context of the larger SLAC community and its relationship with *BABAR* Web. Section 6 describes the use of two standard methods, an online survey and selective interviews, to understand the current perceptions of how the *BABAR* community sees the *BABAR* Web helping it to achieve its goals. Section 7 concludes the paper and indicates areas for further work.

## 2. THE *BABAR* WEB ENVIRONMENT

The *BABAR* experiment is located at SLAC and operated by an international collaboration. Less than half of its approximately 600 participants reside at SLAC at any given time so effective online participation is critical. Consequently the *BABAR* Web site is the main vehicle for timely communication between collaborators; its aims and the associated community are well-defined and explicitly mentioned on the *BABAR* home page (see Fig. 1). The *BABAR* Web is meant to cater to the needs of this community. Even though there are changes to the total *BABAR* population from time to time, the majority of scientists, engineers and administrators are unchanged and fully identified. Further, the review was conducted under the brief of not seeking to change its *modus operandi* since that also was stable and trustworthy. This specificity and full knowledge of the *BABAR* community and its methods contrasts with user communities of other Web sites which may not have such a comprehensive understanding of exactly whom they address, how many, what environments they operate under and their information output and needs.

The longevity and stability of the *BABAR* Web has led its members to assume that there was a common understanding about its environment and how it operates as well as how it should continue to operate in future. Preliminary discussions to scope the current review, in fact, brought out the points of divergence among the constituents of the *BABAR* Web.

The first point to emerge was that such reviews of the Web site would remain essentially incomplete if they focused exclusively on the obvious user community. There are at least two other perspectives that have a direct bearing on the organization and administration of these Web sites. *BABAR* and similar sites operate within the bounds of specific organizations, funded by agencies that demand adherence to policies decided under different priorities. To *BABAR* users, the first priority of the Web site is to provide services that facilitate the tasks vital to the operation of a complex physics experiment. To them, the Web site is a tool. The funding agencies and the hosting organizations typically operate in a bureaucratic and hierarchical manner. For example, the primary funding agency for the *BABAR* experiment is the U.S. Department of Energy (DoE) which may view the *BABAR* Web site as a vehicle for providing a public persona and announcing the experiment's scientific successes. On the other hand, the hosting organization of the *BABAR* Web site (SLAC) is primarily interested in ensuring that the site operates effectively computationally while not providing a mechanism for system intrusion. Security is also a priority of the DoE which specifies policies to be followed by SLAC. These policies were developed by the DoE not only in regard to high-energy physics (HEP) research but stem from its role as custodian of the major U.S. national laboratories and address issues which may not be relevant to SLAC but must be adhered to nonetheless. Consequently, SLAC administrators' perceptions of operational and strategic problems facing *BABAR* Web administration are different from those of the *BABAR* users, as suggested, for example, by the observations listed in section 1.1.

Figure 1 The *BABAR* Web Home Page
(http://www.slac.stanford.edu/BFROOT/)

Additionally, the *BABAR* Web administrators must deal with a specific and different set of problems that arise during the course of managing the users, content, policies, standards and changing technologies. This management of the 'backend' of the *BABAR* Web site is taken for granted by the users and the funding and institutional administrators alike, as 'part of the job'. It is not always apparent to the users and the (SLAC) administrators that there is an inherent tension in matching the expectations of the collaborative participants in a community Web site with the demands of a hierarchical administration.

## 2.1 Population Characteristics

It can be assumed that the *BABAR* experiment is of special interest to two categories of people and organizations: experiment collaborators (participants) and non-collaborators. They are further classified according to their roles and interests, as shown in Table 1. A sizeable fraction of this population, approximately 200, is permanently located at SLAC, but there is a constant flux through short- and medium-term visits by non-U.S. and U.S. collaborators, attendance at seminars, collaboration meetings, and other activities that lead to temporary relocation of different personnel. However, most of the Web site administrators are located at SLAC.

The last column of Table 1 identifies the main informational needs and interests of each corresponding segment of the *BABAR* community. For the sake of completeness, it may be noted that all collaborators will need information about the *BABAR* administration. Also, 'education' has two distinct divisions, secondary schools and colleges/universities, including promotion of summer programs for teachers. In dealing with the funding agencies, the *BABAR* Web also serves at times as a vehicle for public relations, allowing the interested parties to access/download information and reports to generate broader understanding and support for the *BABAR* experiment.

Table 1 Broad Categorization of the *BABAR* Community

| Main Groups | Sub-categories | Approx. numbers | Informational needs & Interest |
|---|---|---|---|
| **Collaborators** | SLAC & *BABAR* Administrators | 10 | Admin |
| | HEP physicists – Physics as priority | 350 | Research |
| | HEP physicists – Computing as priority | 175 | Research |
| | Job seekers within *BABAR* | ? | Employment opportunities |
| | New graduate students | ? | Projects |
| **Non-collaborators** | HEP Physicists | ? | Research results |
| | Funding agencies | <10 | Progress reports, daily operational statistics, public relations |
| | General public | ? | Latest scientific results in non-specialist terms |
| | Education | ? | Education & teacher training |
| | Job seekers outside *BABAR* | ? | Employment opportunities |

**Note**: "?" in the table indicates that it would be difficult to obtain or estimate accurate numbers for that particular group.

Table 2 Hardware and Software Environment of the *BABAR* Web

| Categories | Description | Notes |
|---|---|---|
| **Hardware** | Sun, Intel | SLAC's standard platforms |
| **Operating systems** | Solaris, Linux, Windows NT 4.0 (server and professional), Windows 2000, Windows XP | |
| **Web servers** | Apache | Serves over 90% of *BABAR* Web content |
| | IIS | Used mainly for pages intended for non-collaborator consumption |
| **Databases** | Oracle | Used by many CGI scripts |
| **Web development** | Text editors, Netscape Composer, FrontPage | Must accommodate whatever the collaborators use |
| **CGI scripting languages** | Perl, ASP | |

## 2.2 Hardware and Software Platforms

Since the collaborators and non-collaborators of the *BABAR* experiment are multi-institutional and multi-national, it would be expected that their hardware and software platforms, tools, practices and standards are also equally diverse. Taken together they have implications for how the collaborators create their documents and Web pages. However, rather than go into those details, which will change over time, we describe the software resources and hardware environment at SLAC as available to *BABAR* in Table 2. There is a *BABAR* Web mirror site at Rutherford-Appleton Laboratory (RAL) in England but the narrower focus here is justified on the grounds that the administration of the *BABAR* Web is centralized at SLAC. In the earlier years of the *BABAR* experiment, in the mid-1990s, when connectivity from Europe to the U.S. was more expensive and limited, the mirror played an important role for non-U.S. collaborators. Today the connectivity issues are not as important but the RAL mirror still serves as a backup site when the SLAC Web server is not available.

### 2.3 The *BABAR* Community's Information Output

To cater to the needs of the *BABAR* community as a whole, the *BABAR* collaborators have produced and continue to produce an enormous amount of information. Table 3 summarizes the total information produced according to identifiable categories of information.

The table does not differentiate among the sub-categories of HEP physicists because all of them contribute to the information output. Managing this information content is a complex task. One level of the complexity can be roughly gauged by the number of documents and their sizes in mega- and giga-bytes. This is the 'physical' side of information as expressed in terms of file formats and handling, sorting and searching for information, and archiving. Another aspect of complexity is associated with the hardware platform, in terms of disk and other storage options and network performance. There is a third aspect of complexity which may be termed as 'semantic'. Here, the nature and importance of content, the age and relevance of the information and its maintenance are matters of concern.

Table 3 Information Output

| Information Producer | Category of information | Approx. no. of documents* | Approx. size (MB) |
|---|---|---|---|
| **Administrators** | Documents for user administration | 2,724 | 863 |
| | Scheduling and documenting meetings, workshops, conferences | 7,254 | 5,042 |
| | documents related to job openings | 116 | 2.5 |
| **HEP Physicists (Static content)** | Physics analysis documentation - physics only | 78,289 | 3,396 |
| | Physics analysis- includes computing details | 18,202 | 1,696 |
| | Journal publications | 200 | 250 |
| | Presentations and talks | 350 | 1,800 |
| | Detector operation & maintenance | 106,919 | 6,679 |
| | Computing | 67,655 | 2,296 |
| | Education and general interest (e.g. scientific knowledge and advances) | 20 | 1 |
| | Measures of performance (e.g. for funding) | 20 | 1 |
| | Workbook – "Getting started in *BABAR*", FAQ | 855 | 24.35 |
| | *Doxygen* – hyperlinked documentation of code base library | 63,305 | 1,995 |
| **HEP Physicists (Dynamic content)** | Daily communications (HyperNews) | 150,000 | 2,500 |
| | Detector operation (electronic logbook) | 22,000 | 500 |
| | Detector monitoring (performance graphs) | 100 | 20 |
| | CVSWeb – software management, revision tracking | 50,000 | 30,000 |

\* This number includes all the static documents in a variety of formats, such as HTML, Microsoft Word, PDF, Microsoft PowerPoint, etc.

### 3. THE *BABAR* COMMUNITY AS A VIRTUAL COMMUNITY

The description of the *BABAR* community and its Web environment in the previous section does not really explain its 'community' nature. A community, especially a virtual community in the sense of spreading geographically and over time, exhibits some specific goal-oriented attributes.[2]

• A virtual community shares knowledge that has both <u>explicit</u> and <u>implicit</u> components

• A virtual community requires collaborators to communicate <u>asynchronously</u>

• A virtual community supports the ability of <u>computing</u> to represent information with new <u>tools</u> allowing a broad range of different people to understand complex or conceptual information and <u>participate</u> in exploring it

• A virtual community indicates an organizational (or community) structure that is flexible enough to optimize individual and group performance under new and changing conditions

•A virtual community should create a sense of sharing experience, perspective, support, and trust between people working toward similar goals or solving problems together

The main goal of the *BABAR* Web environment is to enable the *BABAR* community to function at its best utilizing all of its attributes. The Web environment, therefore, needs to be responsive and adaptable to the changing needs of the community, its internal structures and the external environment that impacts the community members. To that end, the *BABAR* Web administration consciously adopted an authoring *modus operandi* that was implemented very early in the site design process with a goal to facilitate efficient participation by all collaborators.

The physical description of the *BABAR* Web in Section 2 and a presumption of a virtual community model need to be defined within the same processes. The authoring *modus operandi* describes how the *BABAR* administrators and collaborators can work on the *BABAR* Web to effectively generate and report information.

### 3.1 *BABAR* Authoring *Modus Operandi*

•**Goal** – To enable occasional, non-professional authors to effectively produce and publish content that is tool independent and meets accessibility standards

•**Method**
 •Minimalist approach to page authoring
 •A minimum of professionally designed graphics;
 •Basic HTML (standardized at Version 4.0);
 •Minimal client-side programming (e.g., JavaScript);
 •Navigation via a standardized *BABAR* wrapper;
 •Infrequent use of Cascading Style Sheets (CSS).

This minimalist approach enables the members to adopt the best practices and tools according to their circumstances and at the same time meet the goal of sharing explicit and implicit knowledge without over-dependence on specific technologies. Other components of the *BABAR* Web address the remaining goals. The HyperNews is a communication platform that is extensively used for publicizing problems and their solutions while CVSWeb allows code sharing. The electronic logbook facilitates the working of the *BABAR* experiment where the collaborators share, asynchronously, the responsibilities of the *BABAR* operations.

Both the *modus operandi* and the organization of the *BABAR* Web thus meet the goals of the virtual community.

### 4. RE-EXAMINATION AND MODELLING OF THE *BABAR* WEB

While the *BABAR* Web *modus operandi* can be useful for defining the community authoring environment and information sharing in a broader sense, the interactions between the creation of information and its consumption/use must be modelled in greater detail. In addition, it is not obvious how the community is affected by SLAC policies and procedures. In order to address these shortcomings, we modelled the 'internals' of the *BABAR* Web with the main emphasis on information and then the interactions between SLAC and the *BABAR* Web. This section deals with 'information modelling' of the *BABAR* Web.

Figure 2 illustrates how information in the *BABAR* Web is created and consumed. This model brings together in an abstract form both the producers and consumers of information. The left branch of the diagram includes all the groups that create (and consume) information. The earlier descriptions in Section 2 can now be seen more clearly in the context of their role in information handling and use. The earlier, physical descriptions can now be properly seen as attributes of different groups, thus implicitly delineating their responsibilities as well.

The right branch of the diagram shows the non-collaborators and their attributes.

It needs to be pointed out that the model does not suggest that these groups are disjoint. In fact, within the *BABAR* community, individuals may "wear more than one hat", e.g., a HEP physicist may also be an administrator.

This model can be further expanded to explore how information 'moves' from the left branch to the right branch or to trace the origins of the information used by, for example, non-collaborating HEP physicists or schools. The model can also be expanded to show more details of the 'physical' output and Web servers and applications currently in use. This is part of an ongoing development of a more comprehensive and, at the same time, simplified model of the *BABAR* Web that might be applied to other Web sites.

### 5. THE *BABAR* WEB AND SLAC

As explained earlier, the *BABAR* Web is partly autonomous but it is operated by SLAC and hence is constrained by the SLAC policies. The diverse groups that form the *BABAR* community bring their own expectations and practices to the collaboration. The *BABAR* Web administrators strive to meet the community expectations, within the limits of what is feasible at SLAC. Innovative suggestions by the users cannot be automatically implemented on the *BABAR* Web in isolation. On the other hand, the *BABAR* computing community is among the most active and the largest user of SLAC computing resources. There is thus a dynamic tension between the *BABAR* computing and SLAC computing communities, resulting sometimes in an 'us vs. them' attitude. Figure 3 clarifies how both SLAC Computing Services (SCS) and the resources required by the *BABAR* Web could ideally interact.

The administrators of SLAC's computing policy and the *BABAR* management are primarily concerned about the implementation of policies, procedures, user administration, providing tools and packages, etc. As Figure 3 illustrates, SLAC's computing policy (the left branch) has several elements that define the overall laboratory environment. The large arrow at the top signifies this influence. The management of the BABAR Web affects the BABAR computing policy that is mainly geared towards satisfying the users' needs in order for them to work fruitfully and smoothly on the BABAR experiment. The users' experience and changing needs lead to demands which at times must 'travel' horizontally across to SLAC. This is expressed as the smaller arrow at the bottom. The whole diagram thus indicates how conflict resolution may be affected.

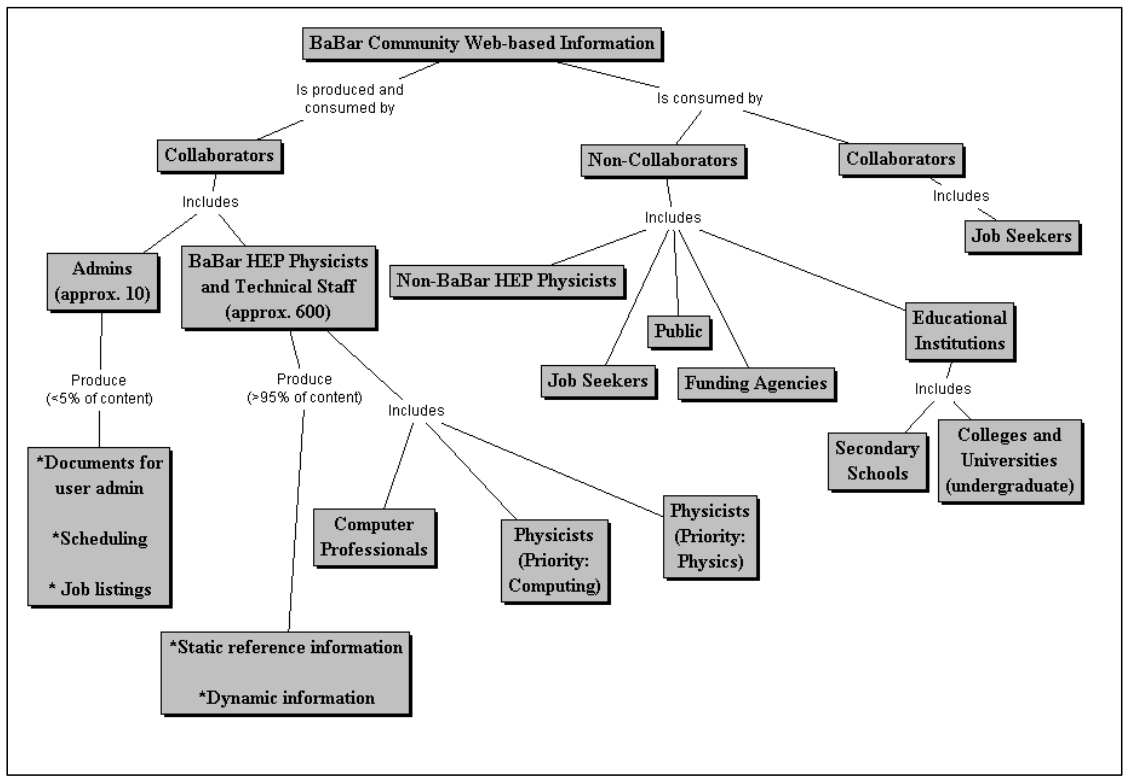

Figure 2 BABAR Web Information Creation and Consumption Model

## 6. THE BABAR WEB REVIEW METHODOLOGY

Sections 2 to 5 have described the scope of the BABAR Web, its informational content, the nature of its community and its relationship with SLAC. This process is a component of "Web site auditing."[3] The second part of the current review has polled the community to elicit the expectations of its users and their actual experience.

Two of the recommended methods, viz. a survey and in-depth interviews of a section of the BABAR community, were used to find out about what the users thought, wanted and the areas where they desired changes and/or improvements. The survey was restricted to only the collaborators.

We also used a third method, viz. Web log analyses, both to understand the usage patterns and check the validity of some the perceptions of the users. Examples of the latter are claims about redundancy or irrelevance of certain pages or of parts of wrappers that standardise navigation throughout the BABAR Web.

The design of the questionnaire in three parts followed the BABAR community model of Figure 1 and Tables 1 and 3. Part 1 asked for general information about the individual members, how they classified themselves primarily (administrators or a specific category of HEP physicists), how often they used the BABAR Web, whether they were responsible for any pages/information themselves and whether they would be willing to be interviewed. Part 2 asked about the collaborators' evaluation of the BABAR Web as a *consumer* of information. The questions related to the usefulness of the features of the BABAR Web (search facility, structure of information, link navigation and accessibility) and about any problems in using these features. The idea of creating a set of pages known as 'core pages' that would be maintained by specific groups was also tested. Part 3 asked similar questions but from the point of the collaborators being *producers* of information. In addition, there were questions about their own responsibilities in creating/maintaining information, technologies used and specific suggestions about solutions to their problems.

The survey was voluntary and its results are being analysed.

The second method was to interview people who were willing to spend time with one of the authors, free-ranging over all the aspects of the *BABAR* Web. These interviews are also being summarized and will be reported in a separate paper.

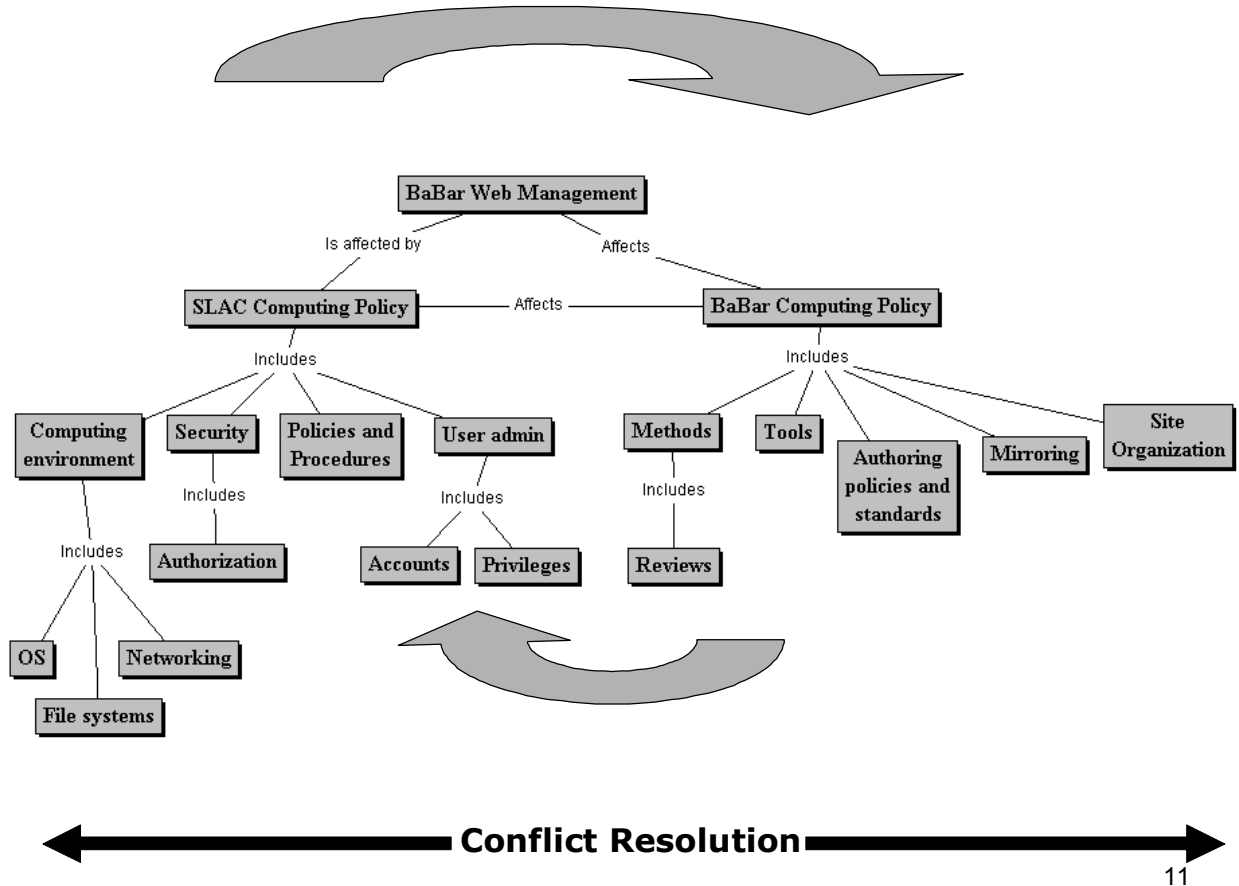

Figure 3  Interaction of *BABAR* Web and SLAC Computing Services

## 7.  CONCLUSIONS AND FUTURE WORK

As a group of scientists and engineers, supported by its own administrators, the *BABAR* community operates semi-autonomously, with international participation.  The community's information needs are generally well-defined and understood.  The output of the *BABAR* experiment continues to grow rapidly, imposing a strain on its resources, its Web site and applications. This evolution is best handled by periodic reviews of the *BABAR* Web environment and suitable changes in practices, technologies and policies that come to the fore as outcomes of the review processes.

The current review has clarified the *BABAR* Web context in SLAC and also as seen by its administrators.  It has led to two more detailed models: a model of the collaborative community (a user model) correlated to information, in terms of the community members being producers and consumers of information (an information model).  It has also clarified several views of information, such as its physical nature (storage and formats), its content (in terms of useful and usable information and the extent of its usage) and its semantic aspects (for searching and structuring including defining the essential or core information).  In themselves, these are not new concepts but they make this and future reviews simpler and more detailed.  Each of these aspects needs to be further elaborated and analyzed.  The usage patterns, for example, will require Web log analyses to be linked to the feedback of the community on what it regards as important and its suggestions for the improvement of the Web environment.

As mentioned earlier, these reviews are also constrained by the SLAC administration policies. The policies and procedures adopted by the *BABAR* administration have to satisfy both SLAC and the *BABAR* community.

The future work will concentrate on reporting on the results of the survey, the interviews and quantitative analyses. Work is also needed in defining and validating metrics useful for community Web sites in general, identifying tools that would be useful to its members and developing the ones that are *BABAR* -specific. Further, the community model presented here can be refined and validated against both within SLAC (other experiments) and outside.

## Acknowledgments

Work supported by Department of Energy contract DE-AC03-76SF00515.